\documentclass[prb,twocolumn,showpacs,preprintnumbers,amsmath,amssymb]{revtex4}
\usepackage{graphicx}
\usepackage{dcolumn}
\usepackage{bm}
\usepackage[T1]{fontenc}
\usepackage[ansinew]{inputenc}
\usepackage{amsmath}
\usepackage{amssymb}
\usepackage{graphicx}
\usepackage{color}
\usepackage[colorlinks]{hyperref}

\begin{document}

\title{Comment on "Protective measurements and Bohm's trajectories" by Y.~Aharonov, B.~G.~Englert, and M.~O.~Scully}

\author{A.~Drezet}
 \affiliation{Institute of Physics, Karl Franzens University, Universit\"atsplatz 5 A-8010
Graz, Austria}

\date{\today}
\begin{abstract}
In the above paper [1], it is claimed that Bohm's mechanics can
 be challenged by a simple gedanken experiment involving a protecting
 measurement. Here we show that this gedanken experiment can be
 justified without problem and without contradicting the axiom attributing a physical
 meaning to the Bohmian trajectorie
\end{abstract}
\maketitle

In a recent publication [1] Aharonov, Englert and Scully  (AES)
claimed having found a challenging dificulty in the interpretation
of Bohm [2] attributing a causal trajectory to any quantum
objects. AES did not criticize the formalism of the Bohm theory
(see also [3]) but only the physical meaning of the calculated
trajectories. Ref.~[1] is the direct continuation of previous
attacks done by these authors [4,5] concerning the
``surrealistic'' nature of the Bohm dynamic. It is shown in [4,5]
that in a ``which-way'' experiment a Bohmian particle can follow a
path completely different from the one intuitively expected. This
conclusion of [4,5] is criticicable since it can indeed be
observed that the strongly non local nature of the Bohm dynamic
prevent us to find a contradiction between the experiment
presented in [4,5] and Bohm's model. This means that the non
intuitive or surrealistic aspect of Bohm's trajectories can not be
experimentally revealed since Bohm's trajectories are changed
after each action of the observers. In the present letter I intend
to show that the new conclusions of [1] are not more challenging
that
the previous arguments of [4,5] and can not be logically accepted.\\
The reasoning of [1] is the following (see [6] for a
generalization): consider a two-particle system composed of two
masses $M$ (the pointer) and $m$ (the observed system) coupled
together by a
 singular potential of the form
 \begin{equation}
H=f\left(t\right)\delta\left(x\right)X,
 \end{equation}
where $X$ and $x$ are the coordinates of $M$ and $m$ respectively,
and $f\left(t\right)$ a  slowly varying function of time (see [1]
for precise definitions). In the regime of very weak perturbation
lasting for a very long time we have  so called protective
measurement which  ``takes advantage of the adiabaticity'' of the
interaction.  Prior to the perturbation  we suppose that the
particle of mass $m$ is bound by an attractive potential in its
ground state being characterized by the energy $E$. Similarly we
suppose the pointer represented by a gaussian wave packet of size
$\delta X$. The effect of the adiabatic perturbation on $m$ is to
change the effective energy $E\left(t\right)$ by a small amount
$\delta E$ proportional to $f\left(t\right)X$.  The essential
effect of the perturbation is to modify the main position and
momentum of the gaussian wave packet. The center of mass is now
shifted by the time dependent interval $\Delta X\left(t\right)$
(equations 15-17 of [1]). This displacement is such that $\Delta
X\left(t\right)\gg \delta X$ for time periods large compared to
the typical duration of interaction and is then unambiguously
observable. Thus$\Delta X\left(t\right)$ is proportional to the
probability $|\Psi(x=0)|^{2}$ of finding the observed
particle at $x=0$ and AES speak for this reason of \emph{protective} measurement for the position.\\
Following AES implies however one paradox: Since the wave function
of the ground state is weakly modified the Bohm trajectories of
$m$ are practically not disturbed by the interaction. Moreover
since the Bohm velocity of $m$ in the ground state is equal to
zero we deduce that the particle stays essentially at the same
place during all the procedure of measurement. This is paradoxical
because the interaction is strongly local and because the pointer
evolves under the influence of a force inducing the displacement
$\Delta X\left(t\right)$. In the words of AES: ``Also under the
circumstances of a protective measurement does the particle
interact locally with another object (the meter) although its Bohm
trajectory does not come anywhere near the interaction region''
[1] and ``...for a Bohmian particle in a given position we can
probe the wave function in most other positions without the
particle never being here''[6]. For AES this challenges every
realistic interpretation of Bohm trajectories and ``...therefore
we can hardly avoid the conclusion that the formally introduced
Bohm trajectories are just mathematical constructions with no
relation
to the actual motion of the particle'' [1]. \\

There is however one hidden assumption in this reasoning. Indeed,
it is implicitly accepted that the process of interaction involved
is a position measurement. Nevertheless the term position
measurement is used generally to describe a von Neumann-like
procedure. We must then compare these two kind of measurements in
order to see if there are not some ambiguities of language. In von
Neumann's measurement theory we can only discuss about position
measurement at $x=0$ if the wave function of $m$ ``collapse'' into
$|x=0\rangle$. This means that the state $|x=0\rangle$ must be
entangled with a well distinguishable state of the pointer.
Consider as an example the position measurement realized when the
system of two particles $m,M$ interact through the potential
\begin{equation}
H'=P_{0}f\left(t\right)D\left(x\right)X ,
 \end{equation}
where $P_{0}$ is a typical momentum of the apparatus and
$D\left(x\right)$ is a function well picked on $x=0$. Additionally
$f(t)$ is the strength of the interaction normalized such that
$\int f\left(t\right)dt=1$ and the interaction is non zero only in
the short interval $[0,\tau]$. In the limit of an impulsive von
Neumann measurement the initial state
$\Psi_{0}(x,X)=\psi(x)\phi(X)$ evolves into
\begin{equation}
\Psi_{1}(x,X)=\psi(x)\phi(X)e^{-iP_{0}D\left(x\right)X}.
\end{equation}
The quantum potential \begin{equation}
Q\left(x,X\right)=-\frac{\hbar^{2}}{2m}\frac{\partial^{2}_{x}|\Psi_{1}(x,X)|}{|\Psi_{1}(x,X)|}
-\frac{\hbar^{2}}{2M}\frac{\partial^{2}_{X}|\Psi_{1}(x,X)|}{|\Psi_{1}(x,X)|}
\end{equation} is not modified by the measurement. However the
dynamics of the particles are defined by the motion laws
$m\ddot{x}\equiv F_{x}=-\partial_{x}U$ and $M\ddot{X}\equiv
F_{X}=-\partial_{X}U$ where $U=Q+H'$. We deduce that both the
measured and measuring system are submitted to an additional force
respectively given by
\begin{equation}
\delta F_{x}=-P_{0}f(t)X\partial_{x}D(x)\simeq
-\frac{P_{0}}{\tau}X\partial_{x}D(x),
\end{equation}
and
\begin{equation}
\delta F_{X}=-P_{0}f(t)\cdot D(x)\simeq -\frac{P_{0}}{\tau}D(x),
\end{equation}
 The future evolution of the joint state will thus be affected
and the Bohm trajectories of $m$ will be in general strongly
modified (this is for example the case if we consider the
Heisenberg microscope experiment). In such a case there is no
problem of interpretation: the Bohm trajectories of $m$ are
locally modified in the region of $x\simeq 0$ and the current
position of the particle is the one which is actually measured.
This is clearly what we could qualitatively expect from a local
and classical theory and it is satisfying that Bohm's theory
conserves, at least, this aspect of the classical
ontology.\\However in [1], despite the fact that the interaction
hamiltonian given by Eq.~1 is local, and that the pointer evolves
in a well distinguishable way, the wave function for m is slightly
modified. It can be then confusing to call by the same name two so
different experimental procedures. We can even ask whether or not
in Bohm's ontological model the weak
measurement of AES can be considered as a position measurement.\\
Moreover for an adherent of Bohm's dynamic there is no problem of
interpretation. In order to see that, we reformulate the analysis
made in [1] and calculate the quantum potential
$Q\left(x,X,t\right)$[2] of the system. In this adiabatic
approximation we deduce
\begin{equation}
Q\left(x,X,t\right)\simeq-\frac{\hbar^{2}}{2m}\frac{\partial^{2}_{x}\psi_{\gamma}(x)}{\psi_{\gamma}(x)}
+ G(X,t)
\end{equation} where $\psi_{\gamma}(x)$ is the adiabatically changed
wave function of the ground state [1], and where the exact
expression of the function $G(X,t)$ is here irrelevant. Using the
definition of $\psi_{\gamma}(x)$ we have additionally
\begin{equation}
-\frac{\hbar^{2}}{2m}\partial^{2}_{x}\psi_{\gamma}(x) +
f\left(t\right)\delta\left(x\right)X\psi_{\gamma}(x) \simeq
E(X,t)\psi_{\gamma}(x),
\end{equation} where the explicit dependence of the energy $E$ is
included. Combining Eqs.~2 and 3 leads to
\begin{equation}
Q\left(x,X,t\right)\simeq -f\left(t\right)\delta\left(x\right)X
+E(X,t)+ G(X,t).
\end{equation}
The total potential affecting the particle $m$ is $U=H+Q$ and we
see that the singular term $-f\left(t\right)\delta\left(x\right)X$
compensates $H$. The force
\begin{equation}
m\ddot{x}\equiv F_{x}=-\partial_{x}U\simeq 0
\end{equation}
 which affects $m$ does not
contain the singular term responsible for the short range
interaction. This means that in spite of Eq.~1 there is no local
interaction acting on the Bohm particle $m$. The main points of
the argumentation of [1] are summarized in this sentence of AES:
``Nevertheless an interaction between the particle and the meter
occurs undoubtedly and its net effect is predictable''. We just
saw that the first point is not true: the guiding wave responsible
for the existence of the quantum potential $Q(x,X,t)$ annihilates
the local effect of Eq.~1 on $m$. However there is effectively an
effect on the pointer $M$ since
\begin{equation} M\ddot{X}\equiv F_{X}=-\partial_{X}U\simeq
-\partial_{X}E\neq 0.
\end{equation} It should be however noted that contrarily to what happens in Eq.~6 $F_{X}$ doesn't come from the local term $-\partial_{X}H$ but
from the term $-\partial_{X}E$ (which is again a pure quantum
effect). These observations justify my doubts concerning the
definition of such an experiment as position measurement. Indeed
since the wave function of the joint state factorize (e.~g.~the
motion of the particle $m$ is not disturbed) we can not know where
the Bohm particle $m$ actually is. Instead of position measurement
it would be better to call the protective measurement a measure of
the density $|\Psi(x\simeq 0)|^{2}$. It is in the context of
Bohm's dynamic a way of measuring the effect of the guiding wave
without involving the particle $m$ itself. The present analysis
proves additionally that we have not the right to say like AES
that ``...the formally introduced Bohm trajectories are just
mathematical constructs with no relation to the actual motion of
the particle''. Indeed the ontology associated with Bohm's theory
involves not only the particle but the guiding wave $\Psi(x,X,t)$
solution of Schr\"{o}dinger equation and responsible for the
existence of the quantum potential $Q(x,X,t)$. Without this term
it will be impossible to understand how the pointer can be
affected by the interaction (e.~g.~if the particle $m$ stays
practically at rest far away from the region $x\simeq 0$). In
other terms the complete analysis of the interaction between the
two bodies $m$ and $M$ requires the presence of a third system:
the guiding wave. We are now able to reconciliate the observations
(iii) and (iv) of the summary of [1], namely the apparent
incompatibility between the existence of a distinguishable
displacement $\Delta X(t)$ of the pointer with the fact that the
``...vast majority of Bohm trajectories of the particle never come
close to the box center where the interaction with the meter
happens''. When we consider the active role of the guiding wave as
a third subsystem interacting with both the measured and measuring
subsystems there is no problem of compatibility between the points
(iii) and (iv) previously mentioned. It is only if we think of
Bohm's theory as a classical dynamic (which means here a dynamic
forgetting the guiding wave) that apparent contradictions appear.
One should observe that the active role of the guiding wave in
Bohm's model has been already pointed out by Hardy [7] and Vaidman
[8] who demonstrated in particular that we can kill a
Schr\"{o}dinger cat with an empty wave. These aspects like the
ones discussed in Refs.~[1,4] can be counter intuitive but they
however do not challenge the reality of particle trajectories in
de Broglie-Bohm's theory. It can be added that the fact that the
quantum potential is not of the short range form expected from a
classical context is not new. It is used for instance to justify
the Aharonov Bohm effect [9,10]. As a conclusion it is worth
mentioning that Bohm's dynamic is self-consistent (at least in the
non relativistic domain) since the probabilistic interpretation
can be justified in its context (with some additional assumptions
[11]). This clearly gives an advantage to Bohm's theory when we
compare it to other interpretations like the one proposed by
Everett [12] which can not justify the randomness of quantum
mechanics [13]. We are free to consider Bohm's trajectories as
physical or not but in both cases there is no way to find a
contradiction between theory and experiments.

\end{document}